\documentclass[conference]{IEEEtran}
\IEEEoverridecommandlockouts
\usepackage{cite}
\usepackage{amsmath,amssymb,amsfonts}
\usepackage{algorithmic}
\usepackage{graphicx}
\usepackage{textcomp}
\usepackage{xcolor}
\usepackage{pax}
\usepackage{url}
\def\BibTeX{{\rm B\kern-.05em{\sc i\kern-.025em b}\kern-.08em
    T\kern-.1667em\lower.7ex\hbox{E}\kern-.125emX}}
\begin{document}

\title{Automated Control Logic Test Case Generation\\using Large Language Models}

\author{\IEEEauthorblockN{
Heiko Koziolek\IEEEauthorrefmark{1}, 
Virendra Ashiwal\IEEEauthorrefmark{1},
Soumyadip Bandyopadhyay\IEEEauthorrefmark{2},
Chandrika K R\IEEEauthorrefmark{2}
}
\IEEEauthorblockA{
	\IEEEauthorrefmark{1}ABB Research, Ladenburg, Germany\\
Email: $<$firstname.lastname$>$@de.abb.com\\
	\IEEEauthorrefmark{2}ABB Research, Bangalore, India\\
Email: $<$firstname.lastname$>$@in.abb.com}
}


\maketitle
\IEEEpubidadjcol

\begin{abstract}
Testing PLC and DCS control logic in industrial automation is laborious and challenging since appropriate test cases are often complex and difficult to formulate. Researchers have previously proposed several automated test case generation approaches for PLC software applying symbolic execution and search-based techniques. Often requiring formal specifications and performing a mechanical analysis of programs, these approaches may uncover specific programming errors but sometimes suffer from state space explosion and cannot process rather informal specifications. We proposed a novel approach for the automatic generation of PLC test cases that queries a Large Language Model (LLM) to synthesize test cases for code provided in a prompt. Experiments with ten open-source function blocks from the OSCAT automation library showed that the approach is fast, easy to use, and can yield test cases with high statement coverage for low-to-medium complex programs. However, we also found that LLM-generated test cases suffer from erroneous assertions in many cases, which still require manual adaption.
\end{abstract}


\section{Introduction}
Control engineering for distributed control systems (DCS) and programmable logic controllers (PLCs) includes designing, implementing, and testing of control logic for complex production processes~\cite{Hollender2010}. Control logic is often formulated in notations such as IEC 61131-3 Structured Text (ST), Function Block Diagrams (FBD), or Ladder Logic (LL)~\cite{Tiegelkamp2010}. Control engineers write programs in these notations manually, which is a laborious process under constant pressure to reduce costs~\cite{Koziolek2020,Koziolek2020b}. Generative AI and Large Language Models (LLM) have gained popularity recently to automate certain development tasks for IT applications~\cite{Fan2023} and could also be utilized for cyclicly executing control software. 

For example, LLMs have been applied in several recent studies for generating test cases for Java and Python code~\cite{Wang2024}, while test case generation for PLC software is still rarely applied in practice. Control engineers often neglect writing proper unit test cases, shifting the testing efforts to the factory acceptance testing phase or commissioning phase, when the software is deployed in its operational environment. Writing appropriate and useful test cases is laborious and may be hard for complex programs. Testing PLC code is challenging due to the cyclic nature of the control logic execution, the internal state retained by the programs, and the importance of timers that interrupt the program flow. 


In practice, several PLC programming environments provide unit testing frameworks, for example, TcUnit\footnote{https://github.com/tcunit/TcUnit} for Beckhoff TwinCAT or CODESYS Test Manager\footnote{https://store.codesys.com/codesys-test-manager.html}. While control engineers can use them to automate test case execution and monitoring, the frameworks still require users to write the test cases manually and not generate them. Researchers have proposed several automated PLC test case generation approaches, often based on model checking~\cite{Simon2015, Bohlender2016}, symbolic execution\cite{Guo2017, Shi2024}, or search-based techniques~\cite{Doganay2013, EbrahimiSalari2023}. These approaches often provide a thorough test case coverage but may suffer from state space explosion problems and produce test cases that are difficult to maintain. No approach has yet attempted to utilize the program interpretation and generation capabilities of LLMs for generating test cases for PLC software.

The contribution of this paper is an initial, automated PLC test case generation approach that queries an LLM to generate test cases. In addition to creating an automated open-source toolchain for test case generation, execution, and reporting, we have conducted initial prompt engineering to increase the quality of the generated test cases with specific instructions. Our method queries an LLM for a table of inputs and expected outputs, which our tooling then converts to executable IEC 61131-3 ST-code. 

We have tested our Test Case Generator on ten open-source function blocks, generated test cases using GPT-4\footnote{https://openai.com/gpt-4}, executed them, gathered statement coverage reports, and checked for succeeding assertions. We found that statement coverage can be achieved with the generated test cases for low to medium-complex function blocks already with simple prompts, but that the LLM-generated assertions are often wrong and require additional manual work. However, the approach may already save control engineers time for formulating test cases and may be combined in the future with other approaches based on symbolic execution or search-based techniques.

The remainder of the paper is structured as follows: Section 2 discusses related work, covering both automated test case generation and LLM-supported engineering. Section 3 motivates and describes our test case generation method before Section 4 summarizes details about the prototypical implementation. Section 5 analyzes the test cases generated for our test samples, interprets the results, and summarizes lessons learned. Section 6 concludes the paper.

\section{Related Work}
\textbf{Automated PLC Test Case Generation:} While there is no research published on LLM-supported test case generation, there are several works that use other methods to synthesize test cases. Doganay et al. \cite{Doganay2013} applied a search-based testing approach on IEC 61131-3 function block diagrams, comparing random testing and hill climbing to increase branch coverage. Jamro \cite{Jamro2015} proposed a dedicated, manually applied test definition language for IEC 61131-3 code to express test scenarios.

Simon et al.~\cite{Simon2015} introduced test case generation for IEC 61131-3 software using a model checker that iteratively created program traces, trying to increase coverage metrics. Transforming the traces into ST-code test programs, they achieved full statement coverage when applying the method on ten function blocks. The approach may suffer from state space explosion for difficult-to-reach statements. Later the approach was enhanced with concolic testing~\cite{Bohlender2016}.

Guo et al.\cite{Guo2017} proposed SymPLC, a method translating IEC 61131-3 code to C-code and then using a symbolic execution tool called Cloud9 to generate test cases. Applied on almost 100 PLC programs, the method produced test cases with high coverage and valid assertions in many cases. Hofer at al.~\cite{Hofer2019} created a native IEC 61131-3 testing library called APTest that supports test case execution on resource-constrained devices.

Grochowski et al.~\cite{Grochowski2022} combined symbolic execution and static analysis to optimize test case generation but found that using summaries of function blocks to speed up the generation process is ineffective. Salari et al.~\cite{EbrahimiSalari2023} translated IEC 61131-3 programs into Python code and generated test cases using the Pynguin test automation framework. Feasibility tests were successful, but the approach still required user intervention. Finally, Shi et al.~\cite{Shi2024} proposed a test generation framework for IEC 61131-3 ST programs that uses dynamic symbolic execution. Experiments showed test cases with high coverage, but the approach may suffer from state space explosion for complex programs.

\textbf{LLM-based unit test generation:} Several researchers have used LLMs for test case generation for procedural, non-PLC software~\cite{Wang2024}. Schaefer et al.~\cite{Schaefer2023} prompted an LLM for JavaScript test case generation and achieved an almost 20\% increase in statement coverage compared to a state-of-the-art feedback-directed test generation approach. For Python code, Bhatia et al.~\cite{Bhatia2023} compared ChatGPT-generated and Pynguin-generated test cases for 109 Python files. They found that statement coverage was nearly identical, but more than 70\% of the ChatGPT-generated assertions failed. Siddiq et al.~\cite{Siddiq2023}: analyzed the quality of LLM-based unit test for Python code and found mixed results regarding statement coverage.

\textbf{LLMs for PLC programming:} LLMs have been used for control engineering, although not yet for test case generation. Koziolek et al.~\cite{Koziolek2023} created a collection of 100 prompts for control logic generation and showed how to generate IEC 61131-3 code from P\&IDs using GPT-4 Vision~\cite{Koziolek2024}. Another approach~\cite{Koziolek2024a} used retrieval-augmented generation to integrate proprietary function blocks into LLM-generated ST-code. Fakih et al. \cite{Fakih2024} used fine-tuned LLMs to generate PLC code and verified its execution using a symbolic model checker. However, none of these approaches applied LLMs for PLC test case generation.

\section{LLM-supported\\PLC Test Case Generation Method}
\subsection{Motivation}
As researchers have proposed several automated testing approaches based on various techniques, LLM-supported PLC test case generation should be viewed as a complementary approach. It can offer special benefits but is best executed in combination with classical methods. Symbolic execution, concolic testing, or model checking are powerful for detecting bugs but also carry special limitations. An LLM-supported test generation approach can benefit from the LLM's capability to understand high-level requirements and context formulated in natural language. It thus can produce test cases aligned with the intended use cases, real-world conditions, and domain-specific constraints. 

Other approaches \cite{Fakih2024} have used LLMs to convert detailed specifications into formal notations for model checking, but in practice, such precise specifications are often unavailable~\cite{Sinha2015}. LLM can work with incomplete or informal specifications for the code and still produce useful test cases. LLMs can potentially also produce more human-like test cases that are easier to understand and maintain, while symbolic execution-based approaches focus on technical paths through the code. While symbolic or concolic testing approaches may suffer from path explosion for complex code, LLMs can be guided to selectively generate test cases for critical and problematic areas of PLC code, possibly based on historic data or risk analysis. LLM-supported test case generation can be performed on the fly and is generally fast (i.e. 10-20 seconds), even for complex code. This may fit well with situations where the requirements are still changing, whereas other approaches may require more effort for regenerating the test cases upon changing requirements.

LLM-generated test cases could support more diverse testing scenarios than concolic testing approaches, which may focus on certain types of errors (e.g., buffer overflows). While concolic testing usually focuses on unit tests (e.g., for individual function blocks), LLM-supported test case generation could be also applied for integration and system testing spanning complex programs. LLM-generated tests may be less powerful than symbolic or concolic testing approaches to uncover specific programming errors since they do not perform a mechanical analysis of the code structures. Therefore, it is advisable to combine LLM-generated test cases with classical test case generation methods.

\subsection{Method Overview}
\begin{figure}[htbp]
\center
  \includegraphics[width=\columnwidth]{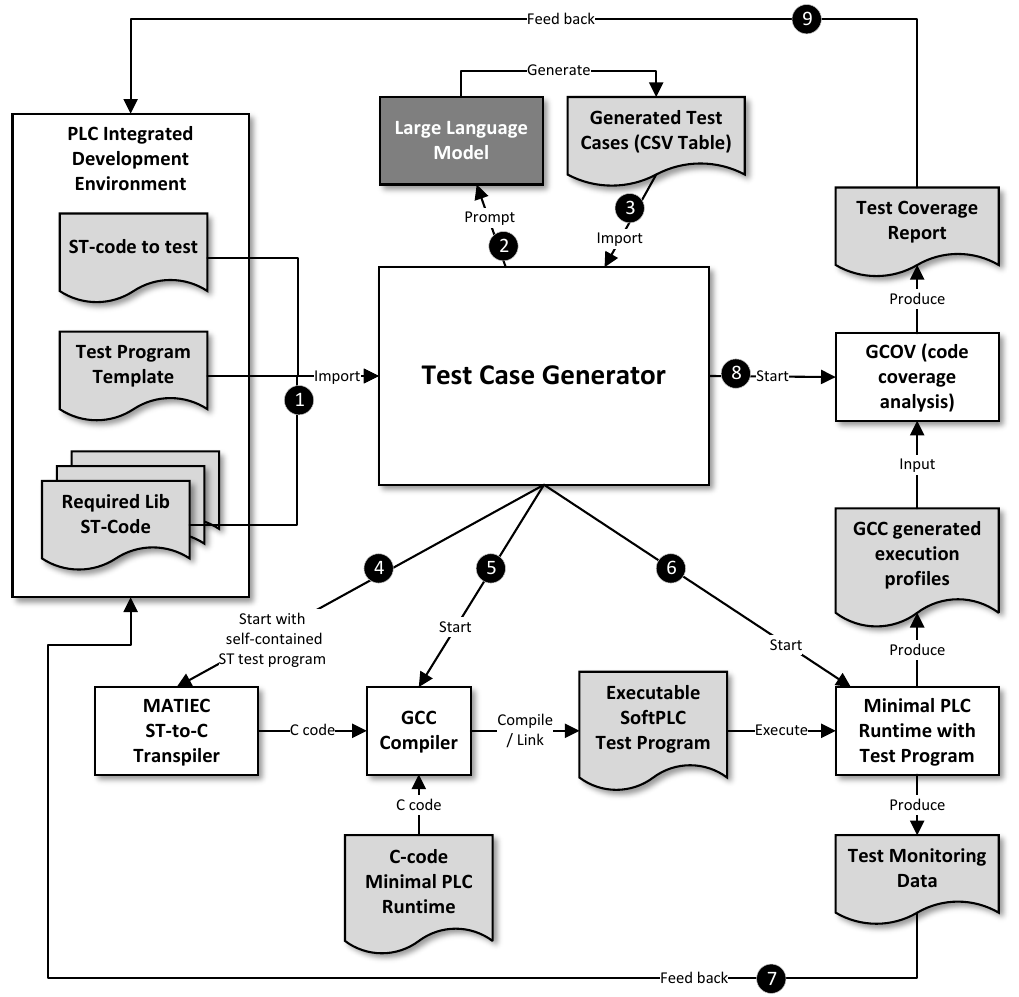}
  \caption{LLM-supported PLC Test Case Generation Approach} 
\label{fig:approach}
\end{figure}

Fig.~\ref{fig:approach} shows an overview of our LLM-supported test case generation method. In step (1), the Test Case Generator takes the ST-code to test (e.g., a function block), a test program template with a task configuration, as well as additional required function blocks for the code to test (e.g., utility functions) as input. The code to test is appended to a prompt for test case generation (detailed later) and sent to an LLM (step 2). The LLM generates the desired test cases as a CSV file containing test states, input values, and expected output values. Using a CSV file instead of directly prompting for ST-code avoids any LLM-caused syntax errors and saves output tokens. The LLM then sends the CSV file back to the Test Case Generator (step 3), which converts it into executable ST-code. Using ST-code to formulate the test cases allows for a simpler execution chain and can improve later maintainability since the test cases can be handled in a PLC IDE side-by-side with the code to be tested. 

In step (4), the Test Case Generator assembles the generated test cases with the test program template and the ST-code of required functions into a self-contained ST-program and calls the MATIEC ST-to-C code transpiler. We chose MATIEC to improve the reproducibility of our method since it is available as open-source. The result of the transpilation is several C-code files and CSV tables storing variable declarations. Combined with the C-code of a minimal PLC runtime (i.e., a simple timer that executes the code cyclicly), the GCC compiler can translate the transpilation result into a self-contained binary that mimics a soft controller (step 5). We call the compiler with flags to enable instrumenting the code for coverage analysis.

The Test Case Generator can then start this binary (step 6), which produces test monitoring data (i.e., a log of the test run cycles and which test cases succeeded or failed), and execution profiles using the GCC instrumentation. The monitoring data can be sent back to the PLC IDE (step 7) for inspection by the control engineer. Finally, the Test Case Generate invokes the GCOV source code coverage analysis tool, which translates the execution profiles into a human-readable test coverage report (step 8). This report is also sent back to the PLC IDE (step 9) so that the user can evaluate the generated test cases.

\subsection{Method Details}
\begin{figure}[htbp]
\center
  \includegraphics[width=\columnwidth]{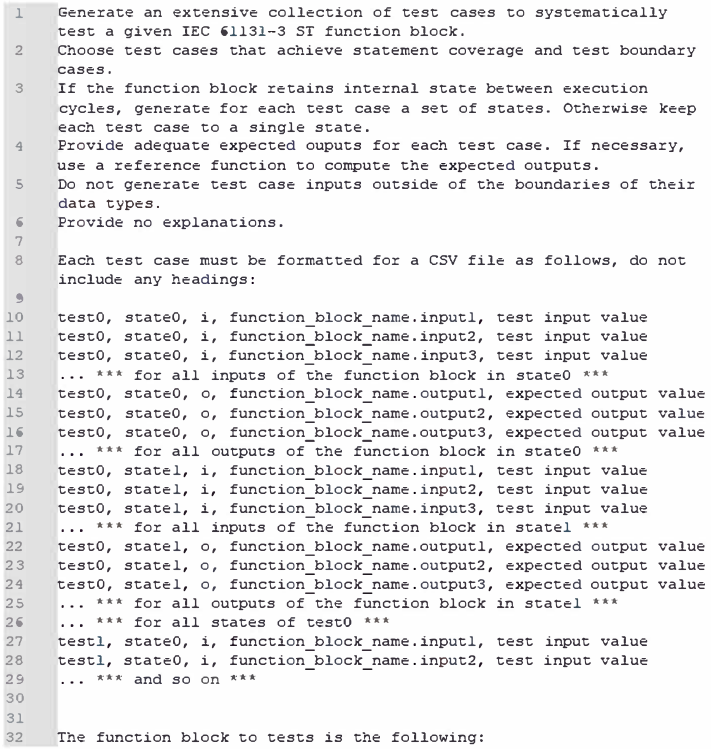}
  \caption{Test Case Generation Prompt} 
\label{fig:prompt}
\end{figure}

A crucial part of our method is the prompt to query LLMs since our method assumes that regular, publicly available LLMs are used without any domain-specific fine-tuning. This makes our method easier to apply and utilize the deep domain knowledge already encoded in the publicly trained LLMs based on their training data but requires some basic prompt engineering. Fig.~\ref{fig:prompt} sketches our prompt template.

Our prompt first provides instructions specifically for test case generation. We ask the LLM to aim for statement coverage of the code, although this is unlikely to be achieved for any non-trivial code. In the second part of the prompt, we provide formatting instructions for the test case as a CSV file. Notice that the parts that shall be adapted by the LLM are marked with "*** instruction ***". Finally, the ST code to test (e.g., a function block) is appended as part three of the prompt.

We assume here first the simple practical case, where the test cases are only generated based on the programmed source code, without further specifications of the intended functionality and semantics. While this limits the LLM in synthesizing correct assertions, it is a common scenario in practice, where a control engineer may not easily provide a precise specification for the programmed code. If such specifications are available they could be appended to the prompt, whose effect needs to be assessed in future work.


\begin{figure}[htbp]
\center
  \includegraphics[width=\columnwidth]{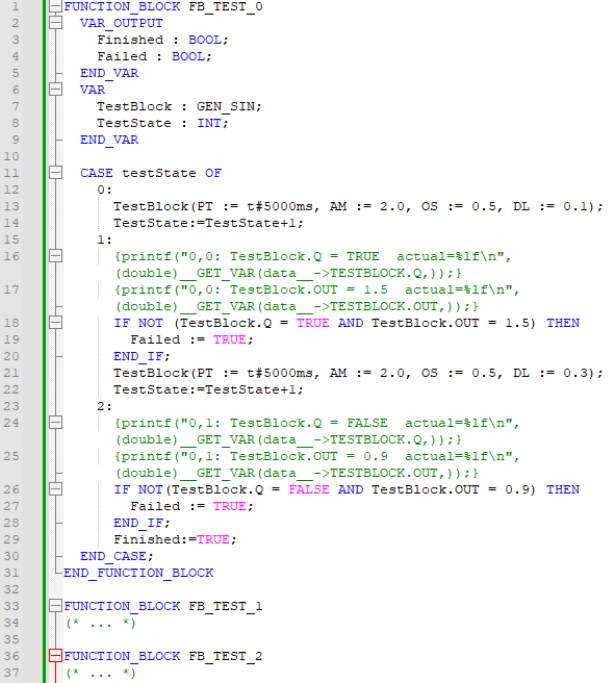}
  \caption{Test Case ST-code} 
\label{fig:test-case}
\end{figure}

In our tests, the prompt from Fig.~\ref{fig:prompt} often resulted in 5-10 test cases from the LLM, which the Test Case Generator then each turned into an ST-code function block as shown in Fig.~\ref{fig:test-case}. The code supports sequential test cases with an arbitrary number of states, so that function blocks retaining an internal state between execution cycles can be appropriately tested. In each state, the function block to test is called with the LLM-generated input variables, and then the expected outputs are checked in the next cycle. The ST-code also contains embedded C-code to later print out test monitoring data during test execution.

\section{Prototypical Implementation}
We implemented a prototype of the Test Case Generator in Python, using LangChain\footnote{https://github.com/langchain-ai/langchain} for LLM interactions. GPT-4-0613 served as the LLM to query and was configured to stream the outputs to standard output so that the test case generation process could be monitored. MATIEC V0.1\footnote{https://github.com/nucleron/matiec} was applied to transpile ST-code to C-code. The resulting C-code for the function block was inserted into a simple soft-PLC execution environment\footnote{https://github.com/Felipeasg/matiec\_examples} and compiled with GCC to a binary. We used the OpenPLC Editor\footnote{https://github.com/thiagoralves/OpenPLC\_Editor} V2.01 to design the test program templates for the testing code. After first creating the test cases directly as ST-code, we changed to creating CSV tables to avoid syntax errors and save tokens. Our Test Case Generator is available\footnote{https://github.com/hkoziolek/LLM-CodeGen-TestGen} on GitHub and can be easily extended for more advanced LangChain features and integrated with a PLC IDE, which is future work.

\section{Experimental Evaluation}

\subsection{Test Setup}
The goal of our experimental evaluation was to test the usefulness of our LLM-supported PLC test case generation method. From this goal, we derived the research question RQ1: ``How good is the quality of the LLM-generated test cases?'' We have not yet attempted to compare the LLM-generated test cases to test cases generated with other methods (e.g., concolic execution). Our method's main contribution is the prompt engineering for LLM instructions that leads to better test cases compared to casual prompting. Therefore, we compare the results from our test case prompt (Fig.~\ref{fig:prompt}, ``enhanced prompt'') to a simplistic prompt that only asks for generating test cases without any specific additional instructions (``simple prompt''). 

As metrics for answering RQ1, we chose the number of test cases generated, the function block statement coverage percentage achieved by executing the generated test cases, and the percentage of successful assertions of the generated function blocks. The latter quantifies what fraction of the generated expected output for the test cases could actually be computed by executing the function blocks. 

As test subjects, we chose 10 ST-code function blocks, which pose different challenges for test case generation. Avoiding proprietary code, we opted for open-source function blocks from the OSCAT Basic Library\footnote{http://oscat.de} and self-written function blocks to aid independent reproduction. The chosen function blocks have at most 200 lines of code but include non-trivial control flows, stateful behavior, and timers, which are typical for PLC code. Tab.~\ref{tab:blocks} provides an overview of these blocks, which have been chosen from different categories, such as logic modules, pulse generators, or mathematics. We tested the blocks only by running the software on workstations without any hardware-in-the-loop.

\begin{table*}[htbp]
\center
  \includegraphics[width=\textwidth]{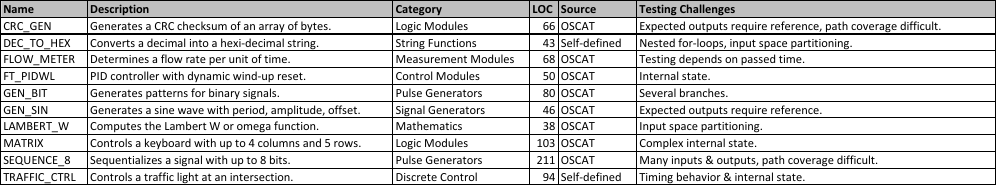}
  \caption{Function Blocks analyzed as test subjects for LLM-based test case generation: multiple categories covered and complimentary test challenges posed} 
\label{tab:blocks}
\end{table*}

\subsection{Result Analysis}
We obtained the source code of our test subject function blocks and applied our LLM-supported test case generation approach that prompted GPT-4 to generate test cases for them. The Test Case Generator also converted each test case from GPT-4 into ST-code according to our method, executed it, recorded the produced outputs, and calculated the achieved statement coverage. Tab.~\ref{tab:results} summarizes the obtained results for the metrics defined for RQ1, including several test cases, statement coverage, and successful assertions both for the simple and enhanced prompt. In the following, we analyze these results by providing details about each function block, starting from the most simple blocks up to the most complex blocks.

\begin{table}[htbp]
\center
  \includegraphics[width=\columnwidth]{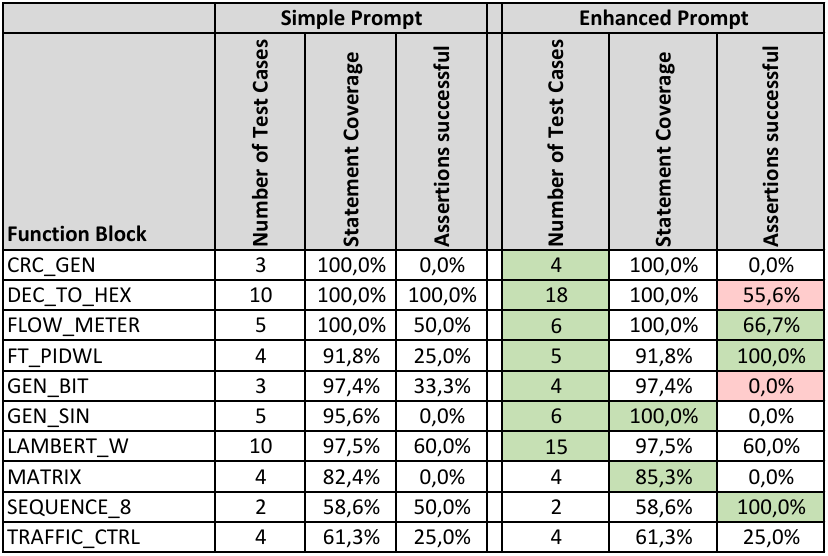}
  \caption{Test case generation results for simple and enhanced prompt: green cells indicate improvements from the enhanced prompt compared to the simple prompt; red cells indicate declines.} 
\label{tab:results}
\end{table}

\begin{figure}[htbp]
\center
  \includegraphics[width=0.9\columnwidth]{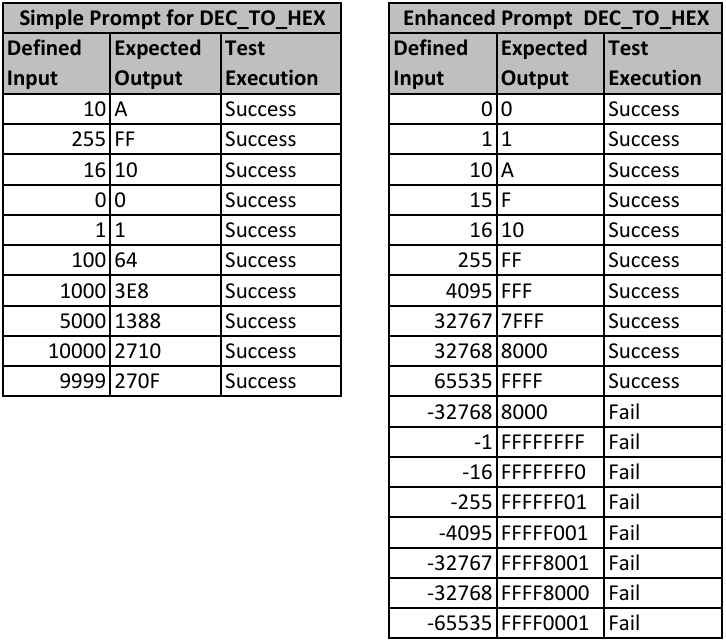}
  \caption{Test cases generated for DEC\_TO\_HEX function block: the enhanced prompt produced more test cases and revealed that the block computes erroneous results for negative inputs.} 
\label{fig:dec-to-hex}
\end{figure}

\textbf{DEC\_TO\_HEX:} This function block converts a decimal input value into a hexadecimal string and was generated using ChatGPT. It has no dependencies on other function blocks and no timing behavior. It contains conditional statements and loops but is comparably easy to test. Both simple and enhanced prompts yielded test cases that achieved 100\% statement coverage, but the enhanced prompt (Fig.~\ref{fig:prompt}) created almost twice as many test cases. For the test cases from the simple prompt, all assertions succeeded, while for the test cases from the enhanced prompt, almost half of the assertions failed. In this case, this was beneficial. Fig.~\ref{fig:dec-to-hex} shows the generated test cases. Test cases yielded by the enhanced prompt correctly revealed that the function block's implementation erroneously handles negative input values, which were not captured by the simple prompt. Furthermore, the test cases of the enhanced prompt cover edge cases where digit boundaries in the output are crossed (e.g., 7FFF and 8000).

\textbf{LAMBERT\_W:} This mathematical OSCAT function block computes the Lambert W function, which is used in mechanical and chemical engineering. It has no time-dependent behavior and computes a single output value. Again, both simple and enhanced prompts yielded test cases with 100\% statement coverage. Both prompts produced seemingly precise expected output floating point values with several decimal places. For the simple prompt, several of these values appear to come from the LLM's training data, which may include mathematical books, while others appear hallucinated. Therefore this function block presents an interesting example of the potential benefits of LLM-based test case generation, where programmers can utilize an LLM's encoded knowledge to quickly produce test cases. For the enhanced prompt that explicitly asked for using reference functions if possible, the LLM actually created a Python code snippet using the Lambert W function in the SciPy math library. The LLM executed this reference function to compute the expected output values. This increased the percentage of successful assertions to 80 percent but still yielded mismatches between the OSCAT function and the SciPy function.

\textbf{CRC\_GEN:} Generating a CRC checksum for an arbitrarily large array of bytes, for this function block it is again easy to achieve high test statement coverage. However, the performed calculations are non-trivial, and therefore an LLM cannot easily generate the expected outputs just by processing the code statistically. The OSCAT documentation provides a link to an online tool\footnote{http://zorc.breitbandkatze.de/crc.html}, which can be manually used as a reference function. In our tests, both the simple and enhanced prompts produced valid inputs, which led to 100 percent statement coverage, but in both cases, the inputs were not chosen in an informed manner, and the expected outputs were completely hallucinated. The LLM was not able to automatically use a reference function in this case. More elaborate prompting would be needed for this function block to generate useful test cases.

\textbf{FT\_PIDWL:} As a PID controller with dynamic wind-up reset from the OSCAT library, this function block retains internal state for computing outputs in the next cycle and also includes timers. The block uses other OSCAT function blocks internally. As our simple prompt only produced simple input/output pairs without considering stateful behavior, only one assertion was successful in a trivial case. Our enhanced prompt produced test cases with multiple states, which allowed first loading up the internal state variables of the function block for calculations in subsequent cycles. This led to all assertions of the enhanced prompt succeeding, albeit it still only covered simple cases. The generated test cases only cover executions with up to three states and no timing behavior. This could be enhanced in future work by including timer behavior in the test programs and elaborating the prompt further.

\textbf{FLOW\_METER:} This function block has a medium test complexity since it requires 5 different inputs as well as time-bound behavior. The block determines a flow rate per unit of time.  The control flow logic is comparably simple, consequently, both the simple and enhanced prompt achieved 100 percent statement coverage. However, due to eight local variables retaining internal state and time-bound behavior, generating correct assertions is challenging. The enhanced prompt achieved one more successful assertion than the simple prompt, but it is unclear if this was caused by the prompt instructions. Testing this block thoroughly requires reflecting its time behavior in the test programs.

\begin{figure}[htbp]
\center
  \includegraphics[width=\columnwidth]{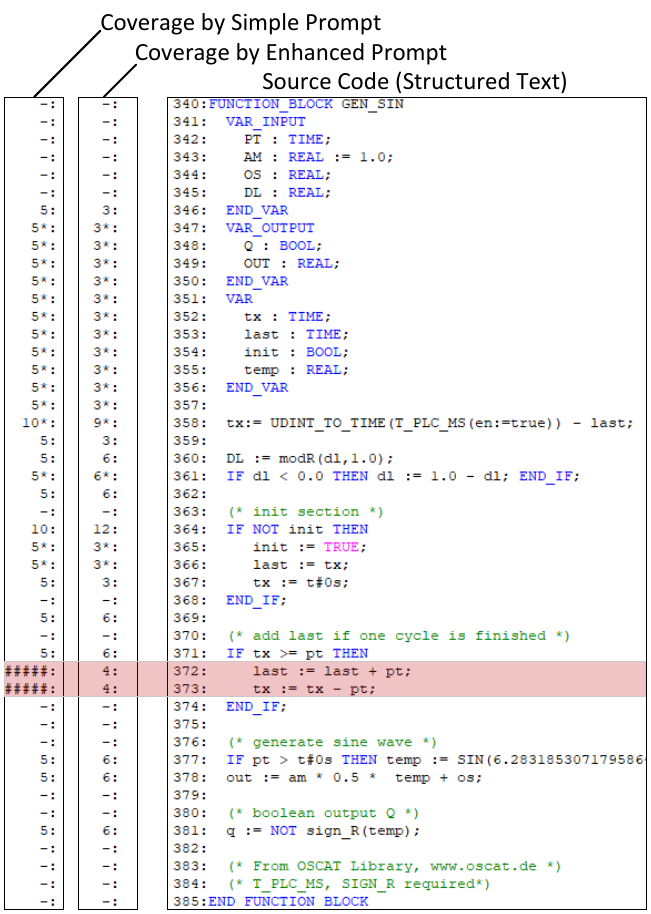}
  \caption{Statement coverage in GEN\_SIN: enhanced prompt yielded coverage of two additional lines of code: 372 \& 373} 
\label{fig:gen-sin}
\end{figure}

\textbf{GEN\_SIN:} Based on the standard sinus function, this function block generates a sine wave with a programmable period. Despite simple control flow logic, the calculated output signal values are time-dependent and therefore difficult for an LLM to estimate. Consequently, in this case, all assertions from the simple and enhanced prompt failed. However, the enhanced prompt yielded a higher statement coverage in this case, which is illustrated in Fig.~\ref{fig:gen-sin}. The simple prompt led to test cases that did not cover line 372 and 373 in the code. The enhanced prompt generated a test case with the input variable PT set to 0.0, which led the execution to covert the respective lines of code in the first cycle. However, it cannot be directly concluded that the instructions for statement coverage in the enhanced prompt caused this since the effect could also result from the inherent randomness of the LLM output.

\textbf{MATRIX:} This function block realizes a matrix controller for a keyboard with a 4x5 button layout. Several conditional statements in the code are dependent on a specific bit encoding given in the function block's specification as well as several cycle executions. Therefore both the test cases for both the simple and enhanced prompt yielded a comparably low statement coverage. The slight improvement achieved by the enhanced prompts appears caused by randomness. The enhanced prompt produced two tests with two states each, which is insufficient to fully cover the code. For this block, none of the generated assertions were successful. In this case, both the simple and enhanced prompts were insufficient to produce high-quality test cases. A more elaborate test case generation procedure is necessary.

\textbf{GEN\_BIT:} This programmable pulse pattern generator includes 8 input variables and 6 output variables, which requires complex test cases. Its outputs are dependent on previous states and, therefore require a sequence of test steps for thorough testing. While both simple and enhanced prompts achieved a high statement coverage, they were unable to produce test cases that hit the code at a conditional statement that compares a previous output with a given input. The enhanced prompt created sequential tests with two states, but could not cover the statements because they included extreme boundary values for the given input. The function block would require test cases with a high number of states and is therefore again challenging to test.

\textbf{TRAFFIC\_CTRL:} This function block implements a traffic light controller with two pedestrian buttons used as input and was generated by ChatGPT. While the logic is intuitive to understand, the block features stateful and timer-dependent behavior. As our generated test programs cannot handle expiring timers yet, the generated test cases both for the simple and enhanced prompt covered only the first of four internal states of the function block. The sequential tests produced by the enhanced prompt included only two subsequent test case states, which were too low to test an entire traffic light sequence. In this case, besides the support for expiring timers in the test program, specific prompts for a single traffic light sequence should be constructed, so that the LLM output tokens could focus on one sequence at a time.

\textbf{SEQUENCE\_8:} With more than 200 lines of code, 27 input variables, 11 output variables, and 16 timers, this 8-bit sequencer OSCAT pulse generator was the most complex function block in our chosen samples. It would require at least 11 test cases for statement coverage. The block contains a long sequence of conditional statements depending on expiring timers. Path coverage of the block would likely require hundreds of test cases and can be considered practically infeasible. As our test programs did not include simulating timer behavior, this function block merely served to explore the LLM capability to generate a large number of input and output values. While the simple prompt generated values for all inputs and outputs, the enhanced prompt only included a selection of the inputs and output without any input timer values and was incomplete. A possible explanation could be that the added instructions in the enhanced prompt consumed too much of the LLM's processing time so it could not follow all the instructions.

\subsection{Result Interpretation \& Lessons Learned}
Our research question RQ1 asked about the quality of the LLM-generated test cases, which we can summarize based on the previously described results. For all test cases, the LLM could generate valid input values, and for most of the function blocks, even the simple prompting led to a high statement coverage already. An instruction for high statement coverage in our enhanced prompt seemed to have little effect on the generated test cases, as for most blocks the statement coverage did not increase significantly.

LLMs seem to be useful for input space partitioning since they can infer these partitions for example by simply recognizing patterns in the included conditional statements. An instruction for creating boundary input values in our enhanced prompt had a visible effect in the generated test cases. This seems plausible since boundary values can be created based on a rather statistical processing of the code as performed by an LLM. Therefore, the generated test cases can save a control engineer time in coming up with such input values manually. Compared to a symbolic testing approach, the results may still be inferior, since in complex cases the statement coverage was far below 100\%.

The generated test cases contained correct assertions only in simple cases, or in cases where the LLM could make use of an existing reference function. Therefore, most of our generated test cases are not directly usable but need manual modifications or further prompting to include correct assertions. However, for simple function blocks, even the simple prompt yielded at least useful input variables. For more complex function blocks with many nested conditionals, stateful behavior, and expiring timers, more sophisticated prompts and test programs are required.

It should be noted that besides the pure test case generation, LLMs showed in our experiments that they can also generate useful explanations of the code under test, design a testing strategy, and even reason on the number of required test cases for statement or path coverage. This feedback could be very valuable for control engineers. 

\section{Threats to Validity}
We analyze several threats that could affect the validity of our results:

\textbf{Internal Validity:} This analyzes whether our experiments established a valid causal relationship between our LLM-supported test case generation and the observed metrics for test case quality. To ensure that all generated test cases were valid, our Test Case Generator turned them into ST-code, which was compiled and executed. We used proven tools, such as MATIEC, GCC, and GCOV to compile the code and calculate coverage metrics. For independent checks of the internal validity, we publish our source code and testing data online. An inherent challenge is the non-deterministic nature of LLM output. Due to the exploratory character of our study, we did not attempt to factor out random outputs by running the queries many times. There is also no statistical significance established for the instructions in our enhanced prompt and the observed improvements, which could be realized in the future. Our test programs cannot yet deal with expiring timers in the ST-code, therefore inherently all assertions relying on timer behavior must fail.

\textbf{Construct Validity:} This analyzes whether our experiments studied the intended concepts with valid constructs. The use case addressed by our Test Case Generator is typical for testing function blocks when building libraries. It does however not cover all kinds of testing scenarios in PLC engineering. We selected typical function blocks as test subjects, which were chosen to cover different kinds of functionality (e.g., mathematical, signal generation, etc.). We used IEC 61131-3 ST-code as a construct for a PLC programming language, which is used in many commercial and open-source PLC IDEs. Other notations, such as ladder logic or function block diagrams could be tested as well, but do not affect the test case generation quality as such. Although we chose non-trivial function blocks from the OSCAT library, there may be more complex blocks in proprietary PLC software, which we have not tested so far. We used popular LLM technology, using GPT-4 as LLM, and LangChain for prototyping, which are both typical constructs in this area.

\textbf{External Validity:} This analyzes whether our study is generalizable to other test subjects or contexts. Our prompts are application and notation-independent and can be applied for test case generation for basically any programs that can be expressed in IEC 61131-3 ST. Even slightly other notations for the function block control logic to test (e.g., vendor-specific flavors of IEC 61131-3) could potentially be used in our LLM-supported test case generation method since LLMs are comparably flexible regarding programming notations used. Our code generator for executing the test cases is currently bound to MATIEC and GCC but could be adapted to other notations if needed. 

\section{Conclusions and Future Work}
We created an automated test case generation approach for IEC 61113-3 function blocks, which utilizes an LLM to generate test case inputs and expected outputs. We designed an enhanced prompt to generate better test cases compared to casual prompting. In our tests, the approach could generate test cases for ten different function blocks in a short amount of time but was still limited by missing support for timers. We learned that LLMs have trouble inferring correct assertions since they only process the code to test statistically. 

Our approach and prototype could eventually lead to refined tooling integrated into PLC development environments that could save control engineers time and effort for creating test cases manually. Researchers get a deeper analysis of the test case generation capabilities of LLMs and can work on novel concepts to address the many open challenges.

As future work, the test case generator can be enhanced with more sophisticated chain-of-thought or tree-of-thought prompting~\cite{Yao2024} to improve the test case generation procedure. It could be investigated if an agent-based approach for the test case generation would be beneficial\cite{Qian2023}. For creating high-quality test cases, the approach could be combined with the classical test case generation approach using symbolic execution~\cite{Shi2024} or search-based techniques~\cite{Doganay2013} into a hybrid approach that combines the benefits of these approaches.

\bibliographystyle{IEEEtran}
\bibliography{etfa2024}

\end{document}